# Intrinsic Cut-off Frequency in Scaled Graphene Transistors

Kartik Ganapathi[†,a)], Youngki Yoon[†,b)] and Sayeef Salahuddin[c)]

Department of Electrical Engineering and Computer Sciences,

University of California, Berkeley, CA, 94720, USA.

## ABSTRACT

Using 2-D self-consistent ballistic quantum transport simulations, we investigate the short-channel behavior of graphene field-effect transistors and its impact on the device transconductance and subsequently the intrinsic cut-off frequency ($f_T$). Although with thin oxides, $f_T$ expectedly scales inversely with the gate length, significant band-to-band tunneling at OFF state leads to a departure from this trend in case of thick oxides. We also examine the effect of achieving better electrostatics at the cost of increased gate capacitance and illustrate that this can indeed degrade the $f_T$. These considerations should be implicit in the optimization of graphene transistors for high-frequency applications.

a) kartik@eecs.berkeley.edu

b) yyoon@eecs.berkeley.edu

c) sayeef@eecs.berkeley.edu

[†] These authors contributed equally to this work.

It is generally agreed that graphene could be a viable candidate for light-weight, flexible, high-frequency, analog applications.[1-4] However, one of the most important factors in limiting the maximum cut-off frequency ($f_T$) in experimentally fabricated long-channel graphene field-effect transistors (GFETs) has been mobility degradation, which has been attributed either to scattering by phonons including those stemming from the substrate[5] or to the relatively low quality of gate oxide.[6] Consequently, experimental measurements of $f_T$ (= $g_m/2\pi C_G$, where $g_m$ and $C_G$ are respectively the transconductance and the total gate capacitance) in these long-channel devices have shown a gate length ($L_G$) dependence of $1/L_G^2$, due to the $1/L_G$ scaling of $g_m$ in the dissipative regime.[6] On the other hand, recent experiments on short-channel GFETs have shown a near $1/L_G$ variation of $f_T$, which has been conjectured to be a signature of contact-limited transport due to gate-length independence of $g_m$ in this regime.[7] Given the importance of $L_G$ scaling for achieving increasingly higher $f_T$, a rigorous understanding of their inter-dependence is critical. While the ultimate RF performance of graphene FETs has been explored previously using numerical simulations [8] and analytical modeling, [9] a detailed understanding of how scaling the gate length will impact $f_T$, especially for short-channel devices, is still missing. In this letter, we address this issue in GFETs using self-consistent ballistic quantum transport simulations by varying gate length between 20 and 200 nm. Our results show that the intrinsic cut-off frequency of scaled GFETs is expected to follow a $1/L_G$ behavior only for very small effective oxide thicknesses (EOTs). By contrast, in case of large EOTs, we observe a considerable departure from this dependence owing to enhanced short-channel behavior. In light of this observation, we study the effect of oxide thickness on $f_T$ for $L_G$ = 20 nm wherein we show that small EOTs, surprisingly, decrease the cut-off frequency of the device due to reduced carrier velocity. Therefore, these implications of short-channel effects need to be carefully considered while optimizing graphene FETs for high-frequency operation.

The structure of the simulated device is shown in Fig. 1. The source and drain contacts are assumed to be metallic with zero Schottky barrier height. The gate length and oxide thickness are varied. Though majority of the reported devices have a gate underlap in excess of 10 nm on each side, we use a 2 nm gate underlap in all our simulations, since the presence of a larger underlap may only alter our results quantitatively through the introduction of parasitic resistances. The interactions with the substrate have been neglected.

We perform ballistic transport simulations within the Non-Equilibrium Green's Function (NEGF) formalism. A tight-binding Hamiltonian in an atomistic $p_z$ orbital basis is used to describe the band structure of graphene. The source and drain self-energies are calculated following the prescription outlined in Ref. 10. A periodic boundary condition is applied along the width of the device, and transverse momentum modes within the first Brillouin zone are summed numerically in calculation of charge densities and current. The electrostatic potential, electron and hole concentrations are computed self-consistently by iteratively solving NEGF equations with Poisson's equation wherein Dirichlet boundary conditions are used for all the terminals. We note that our formalism implicitly takes care of the reduction in effective channel length due to pinch-off behavior at large values of drain bias.

Figures 2(a) and 2(b) show the plots of drain current ($I_D$) and transconductance as a function of gate voltage ($V_G$) at a drain voltage ($V_D$) of 0.4 V for different gate lengths of GFETs with EOT of 0.5 and 25 nm respectively. Evident from the characteristics is the fact that while ON-state ballistic current shows little $L_G$ dependence, the OFF current is increased for shorter gate lengths. The increase is more pronounced in case of larger EOT due to short-channel effects, which will be discussed in detail subsequently. This observation is reflected in the corresponding $g_m$ characteristics as well (on the right axis of Figs. 2(a) and 2(b)), wherein a gate-length-independent ON current and a larger OFF current at shorter gate lengths result in a smaller $g_m$ at all gate voltages. We also note that peak $g_m$ is significantly reduced by varying EOT from 0.5 nm to 25 nm, the consequences of which will be discussed later.

In order to understand the OFF-state degradation for shorter gate lengths, we have plotted energy resolved current, i.e. $I(E)$ (= $T(E)$ × ($f_1(E) - f_2(E)$), where $T(E)$ is the transmission coefficient at a given energy $E$, and $f_1$ and $f_2$ the Fermi-Dirac distributions corresponding to source and drain electrochemical potentials respectively) for different values of $L_G$ with EOT = 25 nm (Fig. 2(c)). The plot shows a predominantly large current for $L_G$ = 20 nm. This stems from a larger lateral electric field, as observed in Fig. 2(d) showing the variation of Dirac point along the device, and hence a significantly enhanced band-to-band tunneling between various transverse mode sub-bands.[11] This is identical to the well-understood phenomena of drain induced barrier thinning in band-to-band tunneling transistors based on conventional semiconductors.

Figures 3(a) and 3(b) depict the variation of peak ballistic $g_m$ with $L_G$ for two different EOTs – 0.5 and 25 nm, respectively. While the peak transconductance saturates at large gate lengths, it drops off from this value as $L_G$ decreases due to short-channel effects, as described previously. The gate length at which this degradation starts and the extent of this fall-off at extremely short gate lengths both depend on the EOT. For EOT = 0.5 nm, the onset of degradation happens at around $L_G$ = 60 nm (indicated by dashed line in Fig. 3(a)) and the reduction at $L_G$ = 20 nm is only about 10% from its saturated value, whereas for a large EOT of 25 nm, the corresponding values are $L_G$ = 120 nm and 36%.

Figure 3(c) shows the dependence of maximum $f_T$ (obtained from the peak $g_m$ values in Figs. 3(a) and 3(b)) on the gate length for EOTs of 0.5 and 25 nm.[12] We note that while the $1/L_G$ dependence mostly holds for EOT = 0.5 nm, in case of EOT = 25 nm there is a considerable departure owing to short-channel behavior of $g_m$ as explained above. In view of this observation, the following comments about the recently reported $1/L_G$ dependence of $f_T$ in Ref. 7 (EOT = 38 nm) are in order:

(i) Irrespective of how the intrinsic $f_T$ scales with $L_G$, a $1/L_G$ dependence can be observed extrinsically if the extrinsic transconductance ($g_m^{extrinsic} = \dfrac{g_m}{1 + g_m R_{parasitic}}$, where $R_{parasitic}$ denotes a gate-length-independent parasitic resistance which might have contributions from contacts and/or gate underlap), is dominated by $R_{parasitic}$ and is hence gate-length independent, which happens under the condition that $R_{parasitic} \gg 1/g_m$. We note that realizing low-resistance Ohmic contacts to graphene is still a matter of active research and the contact resistances are typically larger than those in case of state-of-the-art silicon technology.[13-14] However, with the values of $g_m$ obtained from our simulations (<1 mS/μm for EOT = 25 nm), the aforementioned condition appears less likely to hold in case of contacts whose resistances are characteristically in the range of a few hundreds of Ω-μm (e.g. 600 Ω-μm – Pd/Au,[7] 500 Ω-μm – Ni/Au[14]).

(ii) While all our simulations presented herein correspond to ballistic transport, and the inclusion of the effects of various scattering mechanisms is beyond the scope of this study, we hypothesize that the observation of $1/L_G$ dependence of $f_T$ could be possible even in a scenario where the increase in peak ballistic $g_m$ with increasing $L_G$ due to suppression of short-channel behavior (Figs. 3(a) and 3(b)) can be offset by a decrease in $g_m$ due to scattering such that $g_m^{scattering}$ ($= g_m^{ballistic} \times \frac{\lambda}{\lambda + L_G}$, where $\lambda$ is the scattering mean free path) has little $L_G$ dependence.[15] For example, in case of EOT = 25 nm, this can happen, with $\lambda$ = 90 nm, for $L_G \leq 80$ nm (approximating the peak $g_m$ in Fig. 3(b) by a straight line up to 80 nm and a constant saturated value thenceforth). For the effects of electrostatics and scattering to cancel each other, larger short-channel behavior calls for the presence of greater number of scattering events i.e. smaller value of $\lambda$.

Although it's clear that a smaller EOT is essential for achieving better electrostatics, it needs to be investigated if this translates to a larger $f_T$. Figure 4 shows the variation of $g_m$, $C_G$ and $f_T$ as a function of EOT for $L_G$ = 20 nm, wherein it can be seen that $f_T$ degrades considerably at very small EOTs. This indicates that the rate of increase in $C_G$, due to enhanced oxide capacitance $C_{ox}$ (although limited to some extent by the quantum capacitance of graphene), exceeds that of increase in $g_m$. Physically, this can be explained from the fact that $f_T$ is proportional to the group velocity of carriers ($v$) since $f_T \propto g_m/C_G \propto \partial I_D/\partial Q$ and $I_D \propto Q \times v$, where $Q$ is total charge in the device. Our self-consistent simulations reveal that in case of smaller EOTs, peak $g_m$ is achieved at larger electric fields (inset of Fig. 4(b)), thereby giving rise to population of carriers at higher energy sub-bands of graphene having larger effective mass, and hence a smaller velocity.

To summarize, our simulations show that (i) a $1/L_G$ dependence of intrinsic $f_T$ in scaled GFETs with small EOTs, (ii) significant deviation from this behavior for large EOTs due to enhanced band-to-band tunneling in OFF state, and (iii) degradation of $f_T$ in case of extremely thin gate oxides due to reduced carrier group velocity. These results highlight the role of short-channel effects, which must be considered during optimization of GFETs for high-frequency applications. While our results can be expected to remain qualitatively unaltered at lower temperatures as well due to the temperature independence of short-channel behavior, substrate interactions that are known to create a finite bandgap in graphene can alter the situation considerably.[16] This is because while the roll-off in ballistic $g_m$ at shorter gate lengths could be less prominent, these interactions can induce additional effects like substrate phonon scattering and a more rigorous analysis (including ab-initio methods) is needed in such scenarios. We believe that future studies should involve dissipative transport to gain a clearer understanding of the combined effect of scattering mechanisms and short-channel behavior on the $f_T$ scaling trend in GFETs.

# REFERENCES

1  M. S. Lundstrom, Nature Mater., **10** (8), 566-567 (2011).

2  A. K. Geim and K. S. Novoselov, Nature Mater., **6** (3), 183-191 (2007).

3  P. Avouris, Nano Lett., **10** (11), 4285-4294 (2010).

4  J. S. Moon, D. Curtis, M. Hu et al., IEEE Trans. Electron Dev., **30** (6), 650-652 (2009).

5  I. Meric, N. Baklitskaya, P. Kim and K. L. Shepard, IEDM Tech. Dig., 1-4 (2008).

6  Y.-M. Lin, K. Jenkins, D. Farmer et al., IEDM Tech. Dig., 1-4 (2009).

7  Y. Wu, Y.-M. Lin, A. A. Bol et al., Nature, **472** (7341), 74-78 (2011).

8  J. Chauhan and J. Guo, Nano Res., **4** (6), 571-579 (2011).

9  S. O. Koswatta, A. Valdes-Garcia, M. B. Steiner et al., arXiv:1105.1060v1 [cond-mat.mes-hall] (2011).

10  A. Svizhenko and M. P. Anantram, Phys. Rev. B, **72** (8), 085430 (2005).

11  T. Low, S. Hong, J. Appenzeller, S. Datta and M. S. Lundstrom, IEEE Trans. Electron Dev., **56** (6), 1292-1299 (2009).

12  We emphasize that we have, following Ref. [8], calculated the LC oscillation frequency due to kinetic inductance and found it to be larger than our $f_T$ values, thus validating the quasi-static approximation implicit in our NEGF formalism.

13  F. Xia, V. Perebeinos, Y.-M. Lin et al., Nat. Nanotechnol., **6** (3), 179-184 (2011).

14  K. Nagashio, T. Nishimura, K. Kita and A. Toriumi, IEDM Tech. Dig., 565-568 (2009).

[15] The expression for $g_m^{scattering}$ has been obtained by differentiating with respect to $V_G$ the analogous equation for $I_D$.

[16] S. Y. Zhou, G. -H. Gweon, A. V. Fedorov et al., Nature Mater., **6** (10), 770-775 (2007).

**FIGURE CAPTIONS**

Figure 1. Schematic of the simulated graphene field-effect transistor (GFET).

Figure 2. (a) $I_D$ – $V_G$ characteristics (left axis) at $V_D$ = 0.4 V for $L_G$ = 20 (dashed line), 60 (solid) and 100 nm (dash-dot) for EOT = 0.5 nm. Corresponding $g_m$ – $V_G$ characteristics (right axis) for $L_G$ = 20 (circles), 60 (triangles) and 100 nm (squares). (b) Similar plots as (a) but for GFETs with EOT = 25 nm. (c) $I(E)$ for the same gate lengths for EOT = 25 nm at OFF state (d) The corresponding plots of variation of Dirac point ($E_D$), equivalently the electrostatic potential, along the length of the device. The reference energy ($E_D$=0) denotes the source electrostatic potential and $x$ = 0 the source end of the channel.

Figure 3. Variation of peak $g_m$ with $L_G$ for EOT of (a) 0.5 and (b) 25 nm. (c) Plot of $f_T$ as a function of $1/L_G$ (bottom axis) and $L_G$ (top axis) for the same EOTs. Straight lines are guides for $1/L_G$ dependence.

Figure 4. (a) Peak $g_m$ (left axis) and gate capacitance $C_G$ (right axis) vs. EOT for $L_G$ = 20 nm. (b) Variation of cut-off frequency $f_T$ with EOT at the same gate-length. (Inset) Plot of $E_D$ along the device at gate voltages corresponding to peak $g_m$, for EOTs of 0.5 (solid line), 2 (dash) and 5 nm (dash-dot).

**FIGURES**

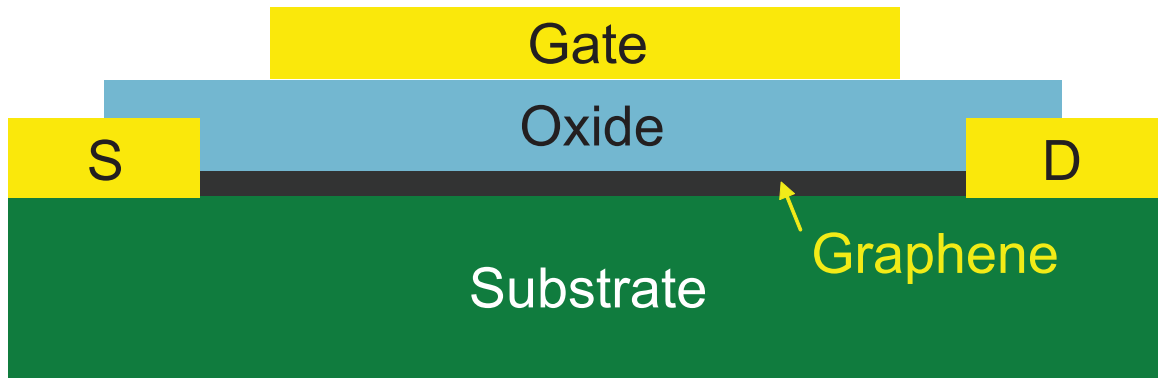

**Figure 1**

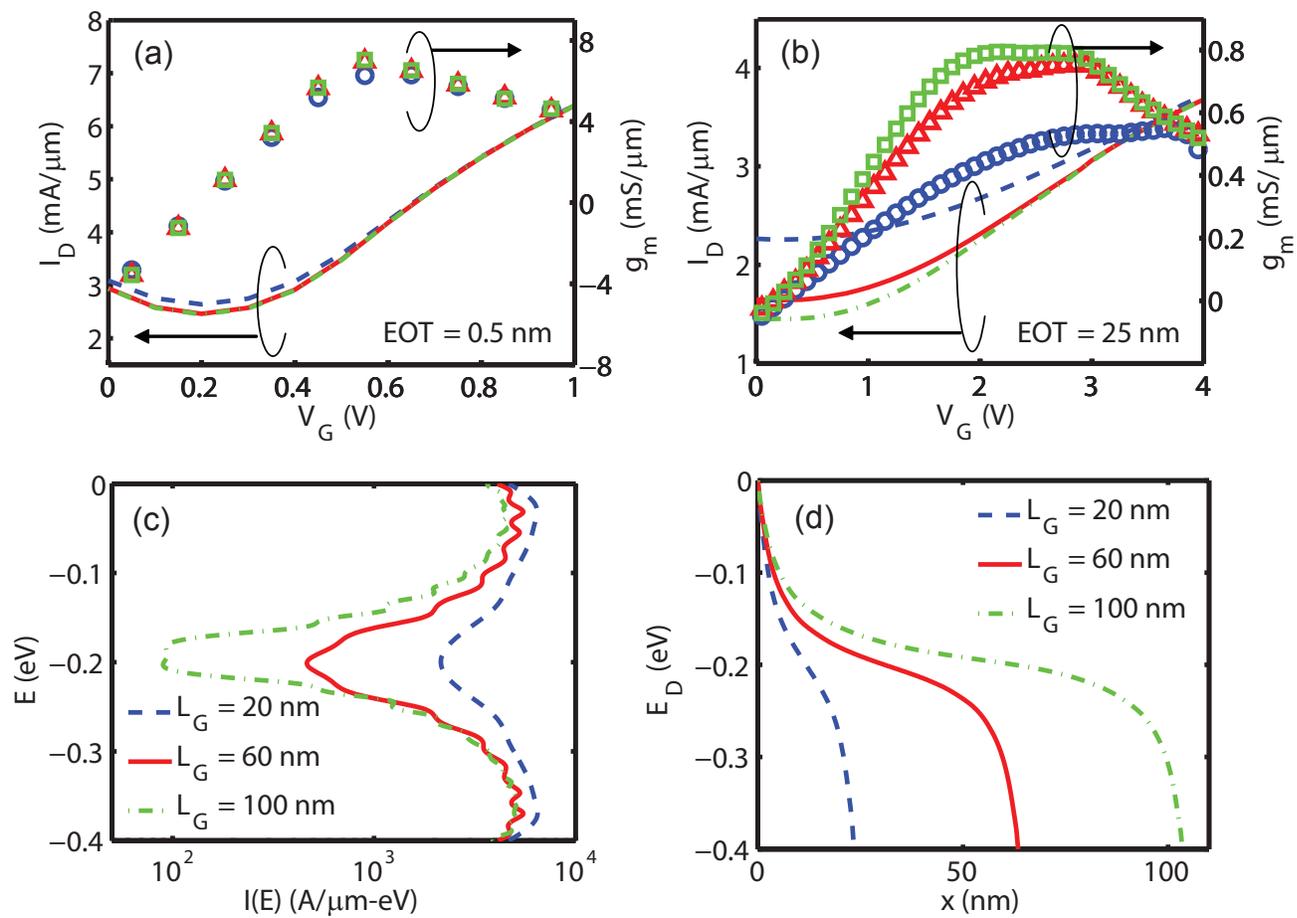

**Figure 2**

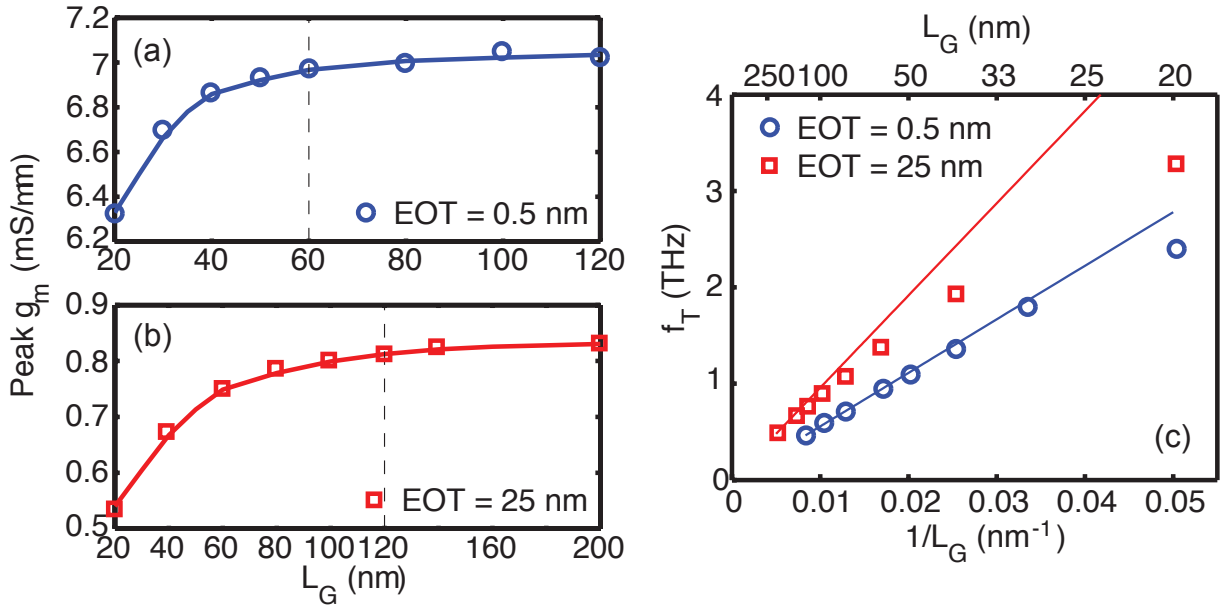

**Figure 3**

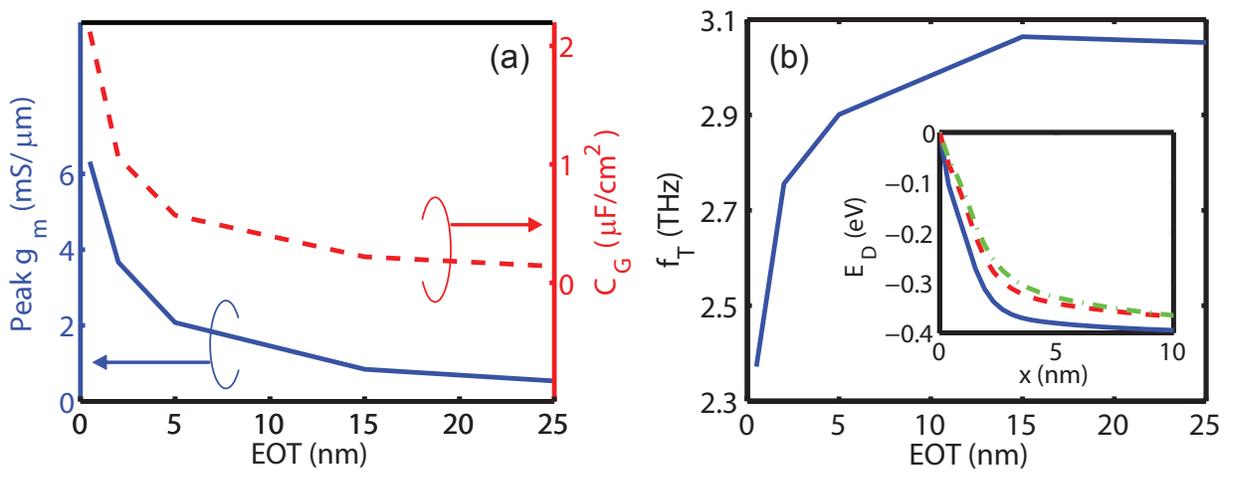

**Figure 4**